\documentclass[12pt,twoside]{article}
\usepackage{fleqn,espcrc1}
\usepackage{epsfig}

\title{Charmonium absorption in the meson-exchange model
\thanks{Supported by NSF Grant PHY-0088934 and Welch Foundation Grant A-1358.}}
\author{Zi-wei Lin\address[TAMU]{
Cyclotron Institute and Physics Department, Texas A\&M University, \\
College Station, Texas 77843-3366} and C. M. Ko\addressmark[TAMU]}

\begin{document}

\maketitle

\begin{abstract}
We review the meson-exchange model for charmonium absorption by hadrons.
This includes the construction of the interaction Lagrangians, the
determination of the coupling constants, the introduction of form factors,
and the predicted cross sections for $J/\psi$ absorption by both
mesons and nucleons. We further discuss the effects due to anomalous 
parity interactions, uncertainties in form factors, constraints 
from chiral symmetry, and the change of charmed meson mass in medium 
on the cross sections for charmonium absorption in hadronic matter.

\end{abstract}

\section{INTRODUCTION}

Suppression of charmonium production in high energy heavy ion collisions 
is one of the most discussed possible signals for the quark-gluon plasma 
(QGP) that is expected to be formed in these collisions \cite{Matsui:1986dk}.
Recent data from the Pb+Pb collision at $P_{\rm lab}=158$ GeV$/c$ in the 
NA50 experiment at CERN-SPS  \cite{Abreu:2000ni} have indeed shown an 
anomalously large $J/\psi$ suppression in events with moderate to large
transverse energy.  Although this anomalous suppression 
may very likely be due to the formation of the QGP \cite{qgp}, 
the conventional mechanism based on $J/\psi$ absorption by comoving 
hadrons has also been shown to contribute significantly to the observed
suppression if the absorption cross sections are taken to be a few
mb \cite{conv}. For heavy ion collisions at RHIC where the number of 
charm mesons per event increases appreciably from SPS, these hadronic 
processes and their inverse reactions may even lead to a net 
production of $J/\psi$ during the hadronic stage of the collisions 
\cite{amptjpsi}. 

Since there is no empirical information on the cross sections 
for charmonium absorption by light hadrons, theoretical models are needed 
to determine their values. These include the perturbative QCD method, 
the quark-interchange model, and the meson-exchange model.  
In the perturbative QCD approach \cite{ope}, charmonia are 
dissociated by the gluons from the colliding light hadron.
The predicted $J/\psi$ dissociation cross section is found to 
increase monotonously with the kinetic energy 
$E_{\rm kin}\equiv \sqrt s-m_h-m_\psi$ but 
has a value of only about $0.1$ mb at $E_{\rm kin}\sim 0.8$ GeV.  
In the quark-interchange model, both the charmonium and the
light hadron are treated as composites of constituent quarks, and the 
absorption of charmonium is via the interchange of the
charm quark in the charmonium with the light quark in the light hadron.  
An earlier study \cite{Martins:1994hd} based on this model gives a 
$J/\psi$ absorption cross section by pion that has a peak value of 
about $7$ mb at the kinetic energy of $\simeq 0.8$ GeV. 
Later studies \cite{Wong:1999zb} predict, however, a peak value of 
only $\sim 1$ mb at the same kinetic energy.  In both cases, 
the cross sections are much greater than that obtained from the 
perturbative QCD method.

In the meson-exchange model, the cross sections between charmonia and hadrons 
are evaluated using effective hadronic Lagrangians derived from
the SU(4) flavor symmetry. In the first application of this model, 
only interaction Lagrangians involving pseudoscalar-pseudoscalar-vector-meson 
couplings are included \cite{Matinyan:1998cb}.  Without employing  
form factors at the interaction vertices, the resulting cross section 
for $J/\psi$ absorption by pion is about 0.3 mb at 
the kinetic energy of 0.8 GeV. In later more complete studies, both 
three-vector-meson and four-point couplings are also included in the 
interaction Lagrangians 
\cite{Haglin:1999xs,Lin:1999ad,Haglin:2000ar,Oh:2000qr,Ivanov:2001th}, 
and much larger cross sections are obtained for $J/\psi$ absorption 
by light hadrons. In contrast to the perturbative QCD approach and 
the quark-interchange model, form factors are needed 
at the interaction vertices in the meson-exchange 
model to take into account the finite size of hadrons. The 
cross sections for $J/\psi$ absorption by light hadrons depend sensitively 
on the form factors. Although there are some studies of the form 
factors involving charmed hadrons \cite{qcd}, the monopole form factors 
with cutoff parameters of about 1 GeV have been used in most meson-exchange
model calculations. The resulting cross sections are a few mb, which
are somewhat larger than those from the quark-interchange model but much 
greater than that from the perturbative QCD method. 

In the following, we shall review in more detail the meson-exchange model 
and its predictions on the cross sections for $J/\psi$ absorption 
by hadrons. We shall also discuss the effects due to anomalous 
parity interactions, uncertainties in form factors, constraints from 
chiral symmetry, and the change of charmed meson mass in medium.

\section{THE MESON-EXCHANGE MODEL}

\subsection{The effective Lagrangian}

To obtain the interaction Lagrangians for charmonia and charmed mesons 
with the least number of free parameters, we start with the free 
Lagrangian for pseudoscalar and vector mesons in the SU(4) limit, i.e., 
\begin{eqnarray}
{\cal L}_0= {\rm Tr} \left ( \partial_\mu P^\dagger \partial^\mu P \right )
-\frac{1}{2} {\rm Tr} \left ( F^\dagger_{\mu \nu} F^{\mu \nu} \right )~,
\label{lagn0}
\end{eqnarray}
where $F_{\mu \nu}=\partial_\mu V_\nu-\partial_\nu V_\mu$, 
and $P$ and $V$ denote, respectively, the properly normalized 
$4\times 4$ pseudoscalar and vector meson matrices in SU(4)
\cite{Matinyan:1998cb,Lin:1999ve}:
\begin{eqnarray} P&=&
\frac{1}{\sqrt 2}\left (
\begin{array}{cccc}
\frac{\pi^0}{\sqrt 2}+\frac{\eta}{\sqrt 6} +\frac{\eta_c}{\sqrt {12}}
& \pi^+ & K^+ & \bar {D^0} \\
\pi^- & -\frac{\pi^0}{\sqrt 2}+\frac{\eta}{\sqrt 6}
+\frac{\eta_c}{\sqrt {12}} & K^0 & D^- \\
K^- & \bar {K^0} & -\sqrt {\frac{2}{3}}\eta 
+\frac{\eta_c}{\sqrt {12}} & D_s^- \\
D^0 & D^+ & D_s^+ & -\frac{3\eta_c}{\sqrt {12}} 
\end{array}
\right ) \;, \nonumber \\[2ex]
V&=&\frac{1}{\sqrt 2}\left (
\begin{array}{cccc}
\frac{\rho^0}{\sqrt 2}+\frac{\omega^\prime}{\sqrt 6}
+\frac{J/\psi}{\sqrt {12}} & \rho^+ & K^{*+} & \bar {D^{*0}} \\
\rho^- & -\frac{\rho^0}{\sqrt 2}+\frac{\omega^\prime}{\sqrt 6} 
+\frac{J/\psi}{\sqrt {12}} & K^{*0} & D^{*-} \\
K^{*-} & \bar {K^{*0}} & -\sqrt {\frac{2}{3}}\omega^\prime
+\frac{J/\psi}{\sqrt {12}} & D_s^{*-} \\
D^{*0} & D^{*+} & D_s^{*+} & -\frac{3J/\psi}{\sqrt {12}}
\end{array}
\right ) \;.  
\label{pv}
\end{eqnarray}

The couplings between pseudoscalar mesons and vector mesons
are obtained through the minimal substitution by replacing 
the partial derivatives with the covariant derivatives, i.e., 
\begin{eqnarray}
\partial_\mu P \rightarrow{\cal D}_\mu P= \partial_\mu P
-\frac{ig}{2} \left [V_\mu P \right ]~,~~~~~~
F_{\mu \nu}\rightarrow\partial_\mu V_\nu-\partial_\nu V_\mu -\frac{ig}{2} 
\left [ V_\mu, V_\nu \right ]~.
\label{ms2}
\end{eqnarray}
The effective Lagrangian is then given by 
\begin{eqnarray}
{\cal L}&=& {\cal L}_0 +\frac{ig}{2} {\rm Tr} 
\left ( \partial^\mu P \left [P^\dagger, V^\dagger_\mu \right ] 
+\partial^\mu P^\dagger \left [P, V_\mu \right ] \right ) 
-\frac{g^2}{4} {\rm Tr} \left ( \left [ P^\dagger, V^\dagger_\mu \right ]
\left [ P, V^\mu \right ] \right ) \nonumber \\
&+&\frac{ig}{2} {\rm Tr} 
\left ( \partial^\mu V^\nu \left [V^\dagger_\mu, V^\dagger_\nu \right ] 
+\partial_\mu V^\dagger_\nu \left [V^\mu, V^\nu \right ] \right ) 
+\frac{g^2}{8} {\rm Tr} 
\left ( \left [V^\mu, V^\nu \right ] 
\left [V^\dagger_\mu, V^\dagger_\nu \right ] \right )~.
\label{lagn1}
\end{eqnarray}
Hermiticity of $P$ and $V$ reduces the above Lagrangian to 
\begin{eqnarray}
{\cal L}&=& {\cal L}_0 + ig {\rm Tr} 
\left ( \partial^\mu P \left [P, V_\mu \right ] \right ) 
-\frac{g^2}{4} {\rm Tr} 
\left ( \left [ P, V_\mu \right ]^2 \right ) \nonumber \\
&+& ig {\rm Tr} \left ( \partial^\mu V^\nu \left [V_\mu, V_\nu \right ] 
\right ) 
+\frac{g^2}{8} {\rm Tr} \left ( \left [V_\mu, V_\nu \right ]^2 \right )~.
\label{lagn2}
\end{eqnarray}
Since the SU(4) symmetry is explicitly broken by hadron masses, terms
involving hadron masses are added to Eq.(\ref{lagn2}) using the
experimentally determined values.
The above effective Lagrangian can also be derived from the chiral 
Lagrangian, which includes both vector and axial vector fields, 
with the axial vector fields removed by gauge transformations 
\cite{Haglin:2000ar,Ivanov:2001th}.

With this effective Lagrangian, various reactions involving 
charmed mesons and $J/\psi$ have been studied. These include  
charmed meson scattering such as $\pi D \leftrightarrow \rho D^*$
\cite{Lin:1999ve,Lin:2000jp} and charmonium absorption
such as $\pi\psi \rightarrow D{\bar D^*}$ 
\cite{Matinyan:1998cb,Haglin:1999xs,Lin:1999ad,Haglin:2000ar,Oh:2000qr,Ivanov:2001th}.
Extension of the effective hadronic Lagrangian to SU(5) flavor symmetry
allows one to study also $\Upsilon$ absorption by hadrons \cite{Lin:2000ke}. 

\begin{figure}
\centerline{\epsfig{file=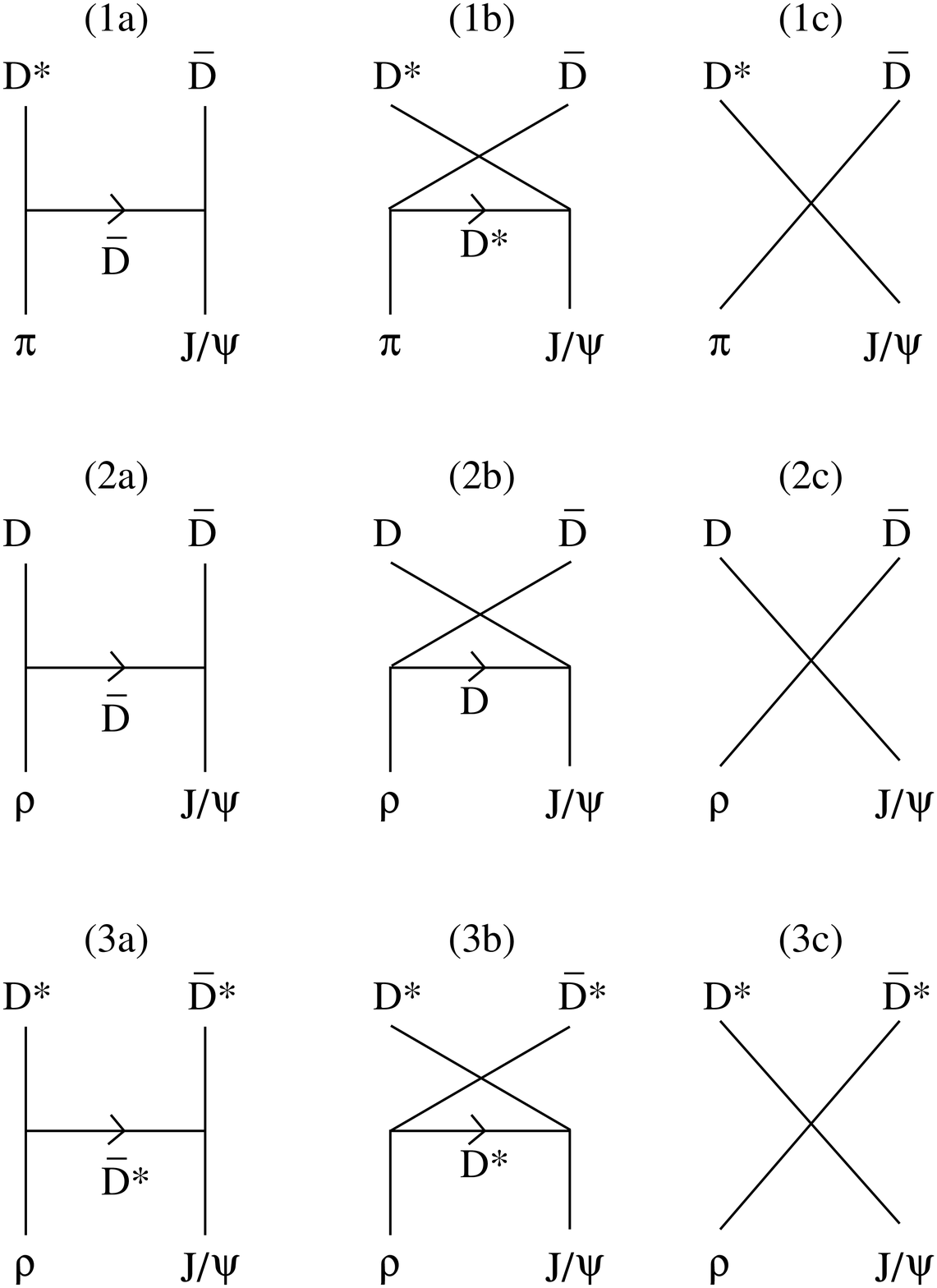,height=3in,width=3in,angle=0}}
\caption{Diagrams for $J/\psi$ absorption by pion and rho meson:   
1) $\pi \psi \rightarrow D^* \bar D$;
2) $\rho \psi \rightarrow D \bar D$; 
and 3) $\rho \psi \rightarrow D^* \bar {D^*}$. Diagrams for
the process $\pi \psi \rightarrow D \bar {D^*}$ are similar to 
(1a)-(1c) but with each particle replaced by its antiparticle.
}
\label{diagrams}
\end{figure}

For $J/\psi$ absorption by $\pi$ and $\rho$ mesons, there are 
following processes:
\begin{eqnarray}
\pi \psi \rightarrow D^* \bar D, ~ \pi \psi \rightarrow D \bar {D^*},~
\rho \psi \rightarrow D \bar D, ~ \rho \psi \rightarrow D^* \bar {D^*}. 
\end{eqnarray}
Their diagrams are shown in Figure~\ref{diagrams} except the process 
$\pi \psi \rightarrow D \bar {D^*}$ which has the same cross section 
as the process $\pi \psi \rightarrow D^* \bar D$. 

As an example, we consider explicitly the first process 
$\pi \psi \rightarrow D^* \bar D$. Its full amplitude, without isospin 
factors and before summing and averaging over external spins, is given by
\begin{eqnarray}
{\cal M}_1 \equiv {\cal M}_1^{\nu \lambda} 
~\epsilon_{2 \nu} \epsilon_{3 \lambda} 
=\left ( \sum_{i=a,b,c} {\cal M}_{1 i}^{\nu \lambda} \right )
\epsilon_{2 \nu} \epsilon_{3 \lambda}, 
\label{m_i}
\end{eqnarray}
with
\begin{eqnarray}
{\cal M}_{1a}^{\nu \lambda}&=& -g_{\pi D D^*} g_{\psi D D}~
(-2p_1+p_3)^\lambda \left (\frac{1}{t-m_D^2} \right ) 
(p_1-p_3+p_4)^\nu, \nonumber \\
{\cal M}_{1b}^{\nu \lambda}&=& g_{\pi D D^*} g_{\psi D^* D^*}~(-p_1-p_4)^\alpha \left ( \frac{1}{u-m_{D^*}^2} \right )
\left [ g_{\alpha \beta}-\frac{(p_1-p_4)_\alpha (p_1-p_4)_\beta}{m_{D^*}^2}
\right ] \nonumber \\
&\times& \left [ (-p_2-p_3)^\beta g^{\nu \lambda}
+(-p_1+p_2+p_4)^\lambda g^{\beta \nu}
+(p_1+p_3-p_4)^\nu g^{\beta \lambda} \right ] , \nonumber \\
{\cal M}_{1c}^{\nu \lambda}&=& -g_{\pi \psi D D^*}~ g^{\nu \lambda}.
\label{pij}
\end{eqnarray}
In the above, $p_j$ denotes the momentum of particle $j$, with 
particles $1$ and $2$ representing initial-state mesons while particles 
$3$ and $4$ representing final-state mesons on the left and right sides 
of the diagrams shown in Figure~\ref{diagrams}, respectively.
The indices $\mu$, $\nu$, $\lambda$, and $\omega$ denote the polarization 
components of external particles while the indices $\alpha$ and $\beta$ 
denote those of the exchanged mesons. 

After averaging (summing) over initial (final) spins 
and including isospin factors, the cross section is given by 
\begin{eqnarray}\label{jpion}
\frac {d\sigma_1}{dt}&=& \frac {1}{96 \pi s p_{i,\rm cm}^2} 
{\cal M}_1^{\nu \lambda} {\cal M}_1^{*\nu^\prime \lambda^\prime}
\left ( g_{\nu \nu^\prime}-\frac{p_{2 \nu} p_{2 \nu^\prime}} {m_2^2} \right )
\left ( g_{\lambda \lambda^\prime}
-\frac{p_{3 \lambda} p_{3 \lambda^\prime}} {m_3^2} \right ),
\end{eqnarray}
with $s=(p_1+p_2)^2$ and $p_{i,\rm cm}$ the momentum of initial-state
mesons in the center-of-mass (c.m.) frame. 

\subsection{Current conservation}

The effective Lagrangian in Eq. (\ref{lagn2}) is generated by minimal 
substitution and is thus equivalent to treating vector mesons as gauge 
particles.  Both $VVV$ and four-point couplings in the Lagrangian 
are thus due to the gauge invariance \cite{Haglin:1999xs}. 
The gauge invariance also leads to current conservation
conditions on the scattering amplitudes
\cite{Haglin:1999xs,Lin:1999ad,Lin:2000jp}.
In the limit of zero vector meson masses, degenerate pseudoscalar
meson masses, and SU(4) invariant coupling constants, these conditions are
\begin{eqnarray}\label{current} 
{\cal M}_n^{\lambda_k \dots \lambda_l}~ p_{j \lambda_j}=0~, 
\end{eqnarray}
with the index $\lambda_j$ denoting the external vector meson $j$
in the process $n$ shown in Figure~\ref{diagrams}. For example, 
we have ${\cal M}_1^{\nu \lambda} p_{3 \lambda}=0$ 
and ${\cal M}_3^{\mu \nu \lambda \omega} p_{2 \nu}=0$. 

If the external vector meson is a member of the diagonal elements in 
the vector meson matrix $V$ in Eq.~(\ref{pv}), such as the rho
meson and $J/\psi$, the above current conservation condition is 
valid even for arbitrary hadron masses and coupling constants, reflecting
the flavor conservation in strong interactions. 

\subsection{Coupling constants}\label{coupling}

In all studies based on the meson-exchange model, the coupling constants 
in the effective Lagrangian are determined as much as possible 
from the empirical information. For example, the coupling constant 
$g_{\pi DD^*}$ is determined from the $D^*$ decay width, 
given by $\Gamma_{D^*\to\pi D}=g_{\pi DD^*}^2 p_f^3/(2\pi m_{D^*}^2)$,
where $p_f$ is the momentum of final particles in the rest frame of $D^*$.
The recently measured width of $\sim 96$ keV from the CLEO experiment 
\cite{Ahmed:2001xc} then gives $g_{\pi DD^*}=5.6$.
However, an old value of $g_{\pi DD^*}=4.4$ \cite{Matinyan:1998cb} 
has been used in most studies.  For other three-point coupling constants 
involving the rho meson or $J/\psi$, the vector meson dominance model 
has been used to relate them to the fine structure constant, i.e., 
\begin{eqnarray}
\frac{\gamma_\psi g_{\psi D D}}{m_\psi^2}=\frac {2}{3}e,~~ 
\frac{\gamma_\rho g_{\rho D D}}{m_\rho^2}
+\frac{\gamma_\omega g_{\omega D D}}{m_\omega^2}=\frac {1}{3}e,~~
-\frac{\gamma_\rho g_{\rho D D}}{m_\rho^2}
+\frac{\gamma_\omega g_{\omega D D}}{m_\omega^2}=-\frac {2}{3}e,
\end{eqnarray} 
where $\gamma_V$ is related to the vector meson partial decay width
to $e^+e^-$, i.e., $\Gamma_{Vee}=\alpha\gamma_V^2/(3 m_V^3)$.   
These relations then give
\begin{eqnarray}
g_{\rho DD}=g_{\rho D^* D^*}=2.52~,~
g_{\psi DD}=g_{\psi D^* D^*}=7.64~.  
\end{eqnarray}

Since there is no empirical information on the four-point coupling constants, 
relations derived from the SU(4) symmetry are thus used to determine 
their values in terms of the three-point coupling constants, i.e., 
\begin{eqnarray}
g_{\pi \psi DD^*}= g_{\pi DD^*} g_{\psi DD}, ~
g_{\rho \psi D D}= 2~g_{\rho DD} g_{\psi DD}, ~
g_{\rho \psi D^* D^*}=g_{\rho D^* D^*} g_{\psi D^* D^*}.  
\end{eqnarray}

\subsection{Form factors}\label{form}

To take into account the composite nature of hadrons, form factors are 
needed at interaction vertices. In most studies, the form factors are 
taken to have the usual monopole form at the three-point $t$ channel 
and $u$ channel vertices, i.e., 
\begin{eqnarray}
f_3=\frac {\Lambda^2}{\Lambda^2+{\bf q}^2},
\end{eqnarray}
where $\Lambda$ is a cutoff parameter, and ${\bf q}^2$ is the squared 
three momentum transfer in the c.m. frame.

The form factor at four-point vertices is less known. In 
Ref. \cite{Lin:1999ad} and later in some other studies, it was taken to 
have the form 
\begin{eqnarray}
f_4=\left ( \frac {\Lambda_1^2}{\Lambda_1^2+{\bar{\bf q}}^2} \right )
\left ( \frac {\Lambda_2^2}{\Lambda_2^2+{\bar{\bf q}}^2}\right ).
\end{eqnarray}
In the above, $\Lambda_1$ and $\Lambda_2$ are the two different cutoff 
parameters at the three-point vertices present in the process with 
the same initial and final particles, and ${\bar{\bf q}}^2$ is 
the average value of the squared three momentum transfers in $t$ and 
$u$ channels.  

For simplicity, the same value has usually been used for all cutoff 
parameters, i.e.,
\begin{eqnarray}
\Lambda_{\pi D D^*}=\Lambda_{\rho DD}=\Lambda_{\rho D^* D^*}
=\Lambda_{\psi DD}=\Lambda_{\psi D^* D^*}\equiv \Lambda, 
\end{eqnarray}
and $\Lambda$ is chosen as either $1$ or $2$ GeV to study the 
uncertainties of the resulting cross sections due to form factors. 

\subsection{Cross sections for charmonium absorption}

\begin{figure}
\centerline{\epsfig{file=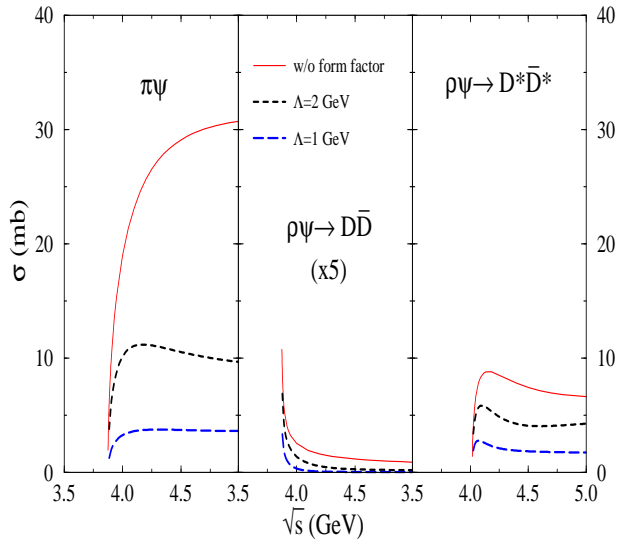,height=3in,width=3in,angle=0}}
\vspace{-0.5cm}
\caption{Cross sections for $J/\psi$ absorption by pion and rho meson 
as functions of c.m. energy with and without form factors.}
\label{sigma-ff}
\end{figure}

Figure~\ref{sigma-ff} shows the cross sections for $J/\psi$ absorption 
by $\pi$ and $\rho$ mesons as functions of the c.m. energy
$\sqrt s$. The cross section $\sigma_{\pi\psi}$
includes contributions from both $\pi \psi \rightarrow D \bar {D^*}$
and $\pi\psi\to D^*{\bar D}$, which have same cross sections.
Solid curves are the results obtained without form factors.
The three $J/\psi$ absorption cross sections are seen to have 
very different energy dependence near the threshold energy.
While $\sigma_{\pi \psi}$ increases monotonously with c.m. energy,  
the cross section for the process $\rho\psi\to D{\bar D}$ decreases 
rapidly with c.m. energy, and that for the process 
$\rho\psi\to D^*{\bar D^*}$ changes little with c.m. energy 
after an initial rapid increase near the threshold.

The results with form factors are shown by the short- and long-dashed curves
for cutoff parameters $\Lambda=2$ and 1 GeV, respectively. 
It is seen that form factors suppress strongly the cross sections 
and thus cause large uncertainties in their values. However, 
the absorption cross sections for $J/\psi$ remain appreciable after 
including form factors at interaction vertices. 
With $\Lambda=1$ GeV, the values for 
$\sigma_{\pi \psi}$ and $\sigma_{\rho \psi}$ are roughly $3$ mb 
and $2$ mb, respectively, and are comparable to those used in 
phenomenological studies of $J/\psi$ absorption by comoving hadrons 
in relativistic heavy ion collisions \cite{conv}.

\section{DISCUSSIONS}

\subsection{Anomalous parity interactions}

There are also anomalous parity interactions which contribute to 
$J/\psi$ absorption by light hadrons \cite{Oh:2000qr}. 
These interactions can be introduced via a gauged Weiss-Zumino 
term in the interaction Lagrangian, and lead to interaction 
Lagrangians of the $PVV$, $PPPV$, and $PVVV$ types such as 
$\pi D^*D^*$, $\rho DD^*$, $DD^*\psi$, $\pi DD\psi$, $\pi D^*D^*\psi$, 
and $\rho DD^*\psi$. 
As examples, we show a few in the following:
\begin{eqnarray}
{\cal L}_{\pi D^*D^*}&=&-g_{\pi D^*D^*}\epsilon^{\mu\nu\alpha\beta}
\partial_\mu D^*_\nu\pi\partial_\alpha {\bar D^*}_\beta,\nonumber\\
{\cal L}_{\pi DD\psi}&=&-ig_{\pi DD\psi}\epsilon^{\mu\nu\alpha\beta}
\psi_\mu\partial_\nu D\partial_\alpha\pi\partial_\beta {\bar D},\nonumber\\
{\cal L}_{\pi D^*D^*\psi}&=&-ig_{\pi D^*D^*\psi}\epsilon^{\mu\nu\alpha\beta}
\psi_\mu D^*_\nu\partial_\alpha\pi {\bar D^*}_\beta
-ih_{\pi D^*D^*\psi}\epsilon^{\mu\nu\alpha\beta}
\partial_\mu\psi_\nu D^*_\alpha\pi {\bar D^*}_\beta,
\end{eqnarray}
where $\epsilon^{\mu\nu\alpha\beta}$ is the totally antisymmetric tensor with
$\epsilon_{0123}=1$. 

As for the coupling constants in the normal interactions, 
the coupling constants in the  anomalous parity interaction Lagrangians 
can mostly be determined from the vector dominance model and the SU(4)
symmetry relations. The coupling constant for $\pi D^*D^*$ can, on the 
other hand, be be related to the $\pi DD^*$ by the heavy quark spin 
symmetry, i.e., $g_{\pi D^*D^*} \sim {\bar M}(D)g_{\pi DD^*}/2$ with 
${\bar M}(D)$ being the average mass of $D$ and $D^*$. 

The anomalous parity interactions give rise to many additional diagrams
for $J/\psi$ absorption. As a result, the cross section for 
absorption by pion is increased by 50\% although that by rho meson
is not much affected \cite{Oh:2000qr}. They also open up new absorption 
processes involving $\eta_c$ and/or $b_1(1235)$ in the final state, i.e., 
$\pi \psi \to\eta_c\rho$ and $\pi \psi \to\eta_c b_1$ via $\omega$ 
exchange \cite{Haglin:2000ar}. The coupling constant  
$g_{\psi\eta_c\omega}$ needed for evaluating these reactions 
can again be determined from the vector 
dominance model, and  
the cross sections are found to be comparable with those processes with 
charmed mesons in the final state.

\subsection{Nuclear absorption}

Since the $J/\psi$ absorption cross sections by pion and rho meson
cannot be directly measured, it is useful to find empirical information 
which can constrain their values. One such constraint is the $J/\psi$ 
absorption cross section by a nucleon, as this process can be viewed as 
the absorption of $J/\psi$ by the virtual pion and rho meson cloud of 
the nucleon \cite{Sibirtsev:2000aw,Liu:2001ce}. From $J/\psi$ production
in photon- and proton-nucleus reactions, the extracted cross section for 
$J/\psi$ absorption by a nucleon is a few mb \cite{Anderson:1976hi}.  

Although the total $J/\psi$ absorption cross section by a nucleon is 
dominated by the process $J/\psi N\to\bar D^*\Lambda_c$ at low 
c.m. energies through charmed meson exchange,  
the process $J/\psi N\to D^*\bar DN$ and $J/\psi N\to \bar D^*DN$ 
due to the virtual pion and rho meson cloud of the nucleon is most important 
at high c.m. energies \cite{Liu:2001ce}. With a cutoff 
parameter of $\Lambda=1$ GeV, the total $J/\psi$ absorption cross section
is at most 5 mb and is consistent with the empirical value.
This result thus indicates that the cross sections for $J/\psi$ 
absorption by pion and rho meson evaluated in previous studies using 
the meson-exchange model are not in contradiction with the empirical 
cross section for $J/\psi$ absorption by a nucleon. 

\subsection{Medium effects}

Since $J/\psi$ absorption by comovers is important in hot and dense
hadronic matter, medium effects on the charmed meson mass 
can affect its cross sections. At finite temperature,
lattice gauge calculations have indicated that 
the linearly rising potential between heavy quarks in 
free space changes to a saturated one as a result 
of the formation of a $\bar Qq-Q\bar q$ pair when their separation 
becomes large \cite{KLP2001}.  
This may be interpreted as the reduction of the charmed meson mass
at finite temperature \cite{DPS,Wong}. 
Although there is no lattice gauge result 
on the heavy quark potential at finite baryon density, explicit calculations 
based on the QCD sum rules analysis \cite{Hayashigaki00,Morath} and 
the quark-meson coupling model \cite{Tsushima99} have shown that 
the charmed meson mass is also reduced at finite density.
With reduced charmed meson mass, not only the threshold for $J/\psi$ 
absorption by comovers is reduced but also the cross sections
for these reactions are enhanced \cite{Sibirtsev:2000aw}.
It is thus important to take into account these medium effects
in studying $J/\psi$ production and suppression in relativistic 
heavy ion collisions. 

\subsection{More on form factors}

Form factors involving charm mesons introduce significant uncertainties 
in the predicted $J/\psi$ absorption cross sections from 
the meson-exchange model. One can get some information on form 
factors from the QCD sum rules. Recent studies based on the three 
point function approach \cite{qcd} show that the $\pi DD^*$ form 
factor for an off-shell pion can be fitted by a Gaussian form 
in the pion four momentum with a range or cutoff parameter 
$\Lambda_\pi=1.2$ GeV, while that for an off-shell $D$ meson is 
best fitted with a monopole form in the charmed meson four momentum 
with a cut-off parameter $\Lambda_D=3.5$ GeV. Since
$\Lambda_D$ is related to $\Lambda$ introduced in Section \ref{form} 
by $\Lambda^2=\Lambda_D^2-m_D^2$. This gives $\Lambda\sim 3$ GeV,
which is in the range of values used for evaluating
the $J/\psi$ absorption cross sections.

Form factors that takes into account the quark substructure of the 
interaction vertices have also been considered \cite{Ivanov:2001th}. 
In this scenario, a three-point interaction vertex is viewed as a 
quark triangle diagram and is associated with a form factor that 
involves the product of three Gaussian wave functions with range 
parameters given by the meson masses. A four-point interaction vertex 
is, on the other hand, viewed as a quark box diagram, and the associated 
form factor is a product of four Gaussian wave function with range 
parameters given again by the masses of the on-shell mesons. 
As a result, the meson-exchange diagrams are suppressed compared 
to the contact interaction diagrams as they involve six Gaussian 
factors. The resulting cross sections with these quark-model 
inspired form factors are thus smaller than those obtained
with the monopole form factors and are comparable to those given by  
the quark-interchange model.

Other types of form factors, ranging from the power law form to the Gaussian 
form, have also been used by other groups \cite{Ivanov:2001th,Oh:2002vg}.
Those results show that the different form factors also affect strongly 
the energy dependence of the $J/\psi$ absorption cross section. 
One can even adjust the form factors to obtain a cross section which 
is very close to that from the quark-interchange model in both its  
magnitude and dependence on the c.m. energy \cite{Wong:1999zb}.

In most calculations, the introduction of form factors has led 
to violation of the current conservation conditions. It is possible
to recover these conditions by introducing more general four-point 
interactions that involve not only the $g_{\mu\nu}$ form but also
all possible lowest-order Lorentz invariant products of the external 
momenta \cite{Haglin:2002ga}. The coefficients in these new amplitudes 
are then adjusted to ensure that the current conservation conditions 
are satisfied.

\subsection{Chiral symmetry}

Although the effective Lagrangian shown in Eq.(\ref{lagn2}) can be derived
from the SU(4) chiral Lagrangian, it involves non-derivative couplings
for pions and thus leads to finite scattering amplitudes when the external
pion four momentum is zero. This violation of the soft pion theorem
is a result of treating charmed mesons on the same footing as
pions. To ensure the SU(2)$\times$SU(2) chiral symmetry, 
the $J/\psi$ absorption cross section by pion has been evaluated
in Ref. \cite{Navarra:2001pz} by dropping those terms in the Lagrangian
that involve non-derivative pion couplings. The resulting cross
section is found to be reduced appreciably. A more consistent way of 
including chiral symmetry in the meson-exchange model is thus needed.

Also, vector mesons are treated as gauge particles in the meson-exchange
model to generate their interactions. Since the SU(4) symmetry is badly 
broken by the heavy quark mass, it is not clear to what extent 
charmonia can be treated as gauge particles. An alternative approach 
\cite{Chan:1996pa} based on both chiral symmetry and heavy quark 
effective theory may provide a more consistent model for the
interactions of charmonia with light hadrons.

\section{SUMMARY}

In summary, we have reviewed the study of $J/\psi$ absorption 
cross sections by $\pi$ and $\rho$ mesons based on the meson-exchange
model. The effective hadronic Lagrangian is generated from the free 
Lagrangian by assuming the SU(4) flavor symmetry and treating
vector mesons as gauged particle.  Using coupling constants
determined either empirically or from relations derived from the
SU(4) symmetry, the resulting cross sections are found to be a few mb
if form factors with reasonable cutoff parameters are introduced at
interaction vertices.  These cross sections are comparable to those from 
the quark-interchange model but are much greater than those from the 
perturbative QCD approach. They are also consistent with that
extracted from $J/\psi$ production in photo- and proton-nucleus 
reactions as the latter can be viewed as $J/\psi$ absorption by the virtual
pion and rho meson cloud of the nucleon. We have also discussed 
the additional contributions from the anomalous parity interactions 
and other anomalous processes involving $\eta_c$ in the final state.
Medium effects due to reduced charmed meson mass are 
mentioned as they not only reduce the threshold of absorption 
processes but also increase their cross sections. We have further
described the attempt of imposing chiral symmetry on pions
by dropping the non-derivative pion couplings in the Lagrangian,
which reduces appreciably the cross section for 
$J/\psi$ absorption by pions. Finally, we have pointed out the 
need to develop an improved approach based on the
chiral symmetry for light hadrons and the heavy quark 
symmetry for heavy hadrons.


\begin{thebibliography}{9}

\bibitem{Matsui:1986dk}
T.~Matsui and H.~Satz,
Phys.\ Lett.\ B 178 (1986) 416.

\bibitem{Abreu:2000ni}
M.C.~Abreu {\it et al.}  [NA50 Collaboration],
Phys.\ Lett.\ B 477 (2000) 28;
{\em ibid} 410 (1997) 337.

\bibitem{qgp}
J.P.~Blaizot and J.Y.~Ollitrault,
Phys.\ Rev.\ Lett.\  77 (1996) 1703;
C.Y.~Wong,
Nucl.\ Phys.\ A 630 (1998) 487.

\bibitem{conv}
W.~Cassing and C.M.~Ko,
Phys.\ Lett.\ B 396 (1997) 39;
W.~Cassing and E.L.~Bratkovskaya,
Nucl.\ Phys.\ A 623 (1997) 570;
N.~Armesto and A.~Capella,
Phys.\ Lett.\ B 430 (1998) 23.

\bibitem{amptjpsi}
B.~Zhang, C.M. Ko, B.A. Li, Z.W. Lin and B.H. Sa,
Phys.\ Rev.\ C 62 (2000) 054905;
B.~Zhang, C.M. Ko, B.A. Li, Z.W. Lin and S. Pal,
{\it ibid.} 65 (2002) 054909.

\bibitem{ope}
D.~Kharzeev and H.~Satz,
Phys.\ Lett.\ B 334 (1994) 155;
F.~Arleo, P.~B.~Gossiaux, T.~Gousset and J.~Aichelin,
Phys.\ Rev.\ D 65 (2002) 014005.

\bibitem{Martins:1994hd}
K.~Martins, D.~Blaschke and E.~Quack,
Phys.\ Rev.\ C 51 (1995) 2723.

\bibitem{Wong:1999zb}
C.Y.~Wong, E.S.~Swanson and T.~Barnes,
Phys.\ Rev.\ C 62 (2000) 045201.

\bibitem{Matinyan:1998cb}
S.G.~Matinyan and B.~M\"uller,
Phys.\ Rev.\ C 58 (1998) 2994.

\bibitem{Haglin:1999xs}
K.L.~Haglin,
Phys.\ Rev.\ C 61 (2000) 031902.

\bibitem{Lin:1999ad}
Z.W.~Lin and C.M.~Ko,
Phys.\ Rev.\ C 62 (2000) 034903.

\bibitem{Haglin:2000ar}
K.L.~Haglin and C.~Gale,
Phys.\ Rev.\ C 63 (2001) 065201.

\bibitem{Oh:2000qr}
Y.~Oh, T.~Song and S.H.~Lee,
Phys.\ Rev.\ C 63 (2001) 034901.

\bibitem{Ivanov:2001th}
V.V.~Ivanov, Y.L.~Kalinovsky, D.~Blaschke and G.R.~Burau,
arXiv:hep-ph/0112354.

\bibitem{qcd}
F.S. Navarra, M. Nielsen, M.E. Bracco, M. Chiapparini and
C.L. Schat, Phys. Lett. B 489 (2000) 319;
F.O. Dur\~aes, F.S. Navarra, M. Nielsen, {\it ibid.}
498 (2001) 169; R.D. Matheus, F.S. Navarra, M. Nielsen and
R. Rodrigues da Silva, {\it ibid.} 541 (2002) 265;
F.S. Navarra, M. Nielsen and M.E. Bracco, Phys. Rev. D 65 (2002) 
037502. 

\bibitem{Lin:1999ve}
Z.W.~Lin, C.M.~Ko and B.~Zhang,
Phys.\ Rev.\ C 61 (2000) 024904.

\bibitem{Lin:2000jp}
Z.W.~Lin, T.G.~Di and C.M.~Ko,
Nucl.\ Phys.\ A 689 (2001) 965.

\bibitem{Lin:2000ke}
Z.W.~Lin and C.M.~Ko,
Phys.\ Lett.\ B 503 (2001) 104.

\bibitem{Ahmed:2001xc}
S.~Ahmed {\it et al.}  [CLEO Collaboration],
Phys.\ Rev.\ Lett.\  87 (2001) 251801.

\bibitem{Sibirtsev:2000aw}
A.~Sibirtsev, K.~Tsushima and A.W.~Thomas,
Phys.\ Rev.\ C 63 (2001) 044906.

\bibitem{Liu:2001ce}
W.~Liu, C.M.~Ko and Z.W.~Lin,
Phys.\ Rev.\ C 65 (2002) 015203.

\bibitem{Anderson:1976hi}
R.L.~Anderson {\it et al.},
Phys.\ Rev.\ Lett.\ 38 (1977) 263.

\bibitem{KLP2001}
F. Karsch, E. Laermann and A. Peikert, 
Nucl. Phys. B 605 (2001) 579.

\bibitem{DPS} 
S. Digal, P. Petreczky and H. Satz,
Phys. Lett. B 514 (2001) 57; Phys. Rev. D 64 (2001) 094015.

\bibitem{Wong} C.Y. Wong,  Phys. Rev. C 65  (2002) 034902;
nucl-th/0112064; C.Y. Wong, T. Barnes, E.S. Swanson and H.W. Crater, 
nucl-th/0112023.

\bibitem{Hayashigaki00} 
A. Hayashigaki, Phys. Lett B 487 (2000) 96.

\bibitem{Morath} 
P. Morath, W. Weise and S.H. Lee, 
{\it  17th Autumn School: Lisbon 1999, QCD; Perturbative or Nonperturbative?}
(Singapore: World Scientific) p 425.

\bibitem{Tsushima99} 
K. Tsushima, D.H. Lu, A.W.Thomas, K. Saito and R.H. Landau, 
Phys. Rev. C 59 (1999) 2824.

\bibitem{Oh:2002vg}
Y.S.~Oh, T.s.~Song, S.H.~Lee and C.Y.~Wong,
arXiv:nucl-th/0205065.

\bibitem{Haglin:2002ga}
K.L.~Haglin,
arXiv:nucl-th/0205049.

\bibitem{Navarra:2001pz}
F.S.~Navarra, M.~Nielsen and M.R.~Robilotta,
Phys.\ Rev.\ C 64 (2001) 021901.

\bibitem{Chan:1996pa}
L.H.~Chan,
Phys.\ Rev.\ D 55 (1997) 5362.

\end{thebibliography}
\end{document}